\documentclass[english,aps,tightenlines,showpacs, showkeys, notitlepage]{revtex4-2}
\usepackage[T1]{fontenc}
\usepackage[latin9]{inputenc}
\usepackage{color}
\usepackage{babel}
\usepackage{amstext}
\usepackage{amssymb}
\usepackage{graphicx}
\usepackage{setspace}
\usepackage{esint}
\usepackage[unicode=true,pdfusetitle,
 bookmarks=true,bookmarksnumbered=false,bookmarksopen=false,
 breaklinks=false,pdfborder={0 0 0},pdfborderstyle={},backref=false,colorlinks=true]
 {hyperref}
\begin{document}
\title{Born--Infeld-$f(R)$ black holes}
\author{Salih Kibaro\u{g}lu$^{1}$}
\email{salihkibaroglu@maltepe.edu.tr}

\date{\today}
\begin{abstract}
We explore black hole solutions in the context of Born-Infeld-$f\left(R\right)$
gravity, a modified gravitational framework that extends both Born-Infeld
and $f\left(R\right)$ theories. By adopting a static, spherically
symmetric spacetime ansatz, we derive an exact black hole solution
and investigate its geometrical structure. We proceed to analyze the
thermodynamic properties of the solution, including the Hawking temperature,
entropy, and specific heat, with particular emphasis on their dependence
on the model parameters. Our results reveal novel thermodynamic behavior
that deviates significantly from the standard predictions of general
relativity. A comparative study with the Schwarzschild-AdS black holes
is also presented, showing how Born-Infeld-$f\left(R\right)$ corrections
alter black hole thermodynamics. 
\end{abstract}
\affiliation{$^{1}$Maltepe University, Faculty of Engineering and Natural Sciences,
34857, Istanbul, Türkiye}
\maketitle

\section{Introduction}

The study of black hole solutions within modified theories of gravity
has gained substantial interest in recent years, largely due to their
potential to address foundational problems in cosmology, high-energy
physics, and quantum gravity. Among the prominent alternatives to
general relativity (GR), Born-Infeld (BI) gravity \citep{born1934foundations,deser1998born,eddington1923mathematical,vollick2004palatini,vollick2005born,Banados2010eddington}
stands out as a compelling candidate. Originally motivated by nonlinear
electrodynamics as a means to regularize field singularities \citep{born1934foundations},
the gravitational extension of the BI action introduces a square-root
determinant structure that naturally incorporates higher-order curvature
corrections. This feature enables the theoretical avoidance of curvature
singularities, such as those appearing in black hole interiors \citep{deser1998born}. 

However, the inclusion of higher-order curvature terms often leads
to the emergence of ghost modes---unphysical degrees of freedom that
signal instability at the quantum level. To circumvent this problem,
the Palatini formulation of BI gravity---drawing inspiration from
Eddington\textquoteright s purely affine gravity framework \citep{eddington1923mathematical}
has been developed. In this approach, the metric and connection are
treated as independent variables, effectively avoiding higher-order
derivatives in the field equations and thereby eliminating the ghost
problem \citep{vollick2004palatini,vollick2005born,Banados2010eddington}.
This formulation is widely referred to as Eddington-inspired Born-Infeld
(EiBI) gravity. EiBI gravity has proven to be a fertile ground for
addressing gravitational phenomena beyond the scope of GR. It has
been successfully applied to early-universe cosmology \citep{Avelino:2012BIcosmologies,Escamilla-Rivera:2012Tensor,Cho:2012Universe,Scargill:2012Cosmology,Kruglov:2013Modified,Yang:2013Linear,Du:2014Large,Kim:2014Origin},
astrophysical modeling \citep{Avelino:2012BIcosmologies}, black hole
physics \citep{Olmo:2014GeonicBH,Sotani:2014Properties,Avelino:2016InnerStructure,Bambi:2016BlackHole},
and wormhole configurations \citep{Lobo:2014Microscopic,Harko:2015Wormhole,Shaikh:2015LorentzianWormhole},
showcasing its versatility and robustness as a modified gravity framework.

Complementing this approach, $f\left(R\right)$ gravity constitutes
a well-established generalization of GR, where the Ricci scalar in
the Einstein-Hilbert action is replaced with a more general function
$f\left(R\right)$ \citep{nojiri2011unified,nojiri2007introduction,Nojiri:2017ncd}.
This modification has been extensively studied for its implications
in late-time cosmic acceleration, dark energy modelling, and high-energy
gravitational phenomena. The unification of EiBI gravity with $f\left(R\right)$
extensions, resulting in BI-$f\left(R\right)$ gravity, yields a rich
theoretical structure that accommodates both nonlinear geometric deformations
and deviations from Einsteinian dynamics \citep{makarenko2014born}.
This model allows for a more comprehensive description of gravitational
phenomena which holds promise for resolving some of the outstanding
problems in modern cosmology and gravitational theory \citep{makarenko2014born,Makarenko:2014nca,odintsov2014born,Elizalde:2016vsd,Kibaroglu:2023BIdS,Kibaroglu:2023CosmBI,Kibaroglu:2024UnimodularBI,Kibaroglu:2025AnisotropicBI}
(for a detailed review, see \citep{jimenez2018born}). This presents
new avenues for exploring the physics of black holes and cosmological
evolution in more general settings.

In this work, we aim to explore static, spherically symmetric black
hole solutions within the BI-$f\left(R\right)$ framework, focusing
on configurations that asymptotically approach Schwarzschild--(Anti)de
Sitter (Sch-AdS) geometries. These solutions are not only important
for understanding the behavior of black holes in curved cosmological
backgrounds but are also of interest in the context of holography
and the AdS/CFT correspondence. The presence of a cosmological constant
or effective curvature scale in such solutions provides a testing
ground for probing gravitational modifications beyond Einstein gravity.

The paper is organized as follows: In Section \ref{sec:Born-Infeld--gravity},
we provide a brief review of the BI-$f\left(R\right)$ gravity theory,
outlining its fundamental principles and relevant field equations.
In Section \ref{sec:Black-Hole-Solution}, we derive a static, spherically
symmetric black hole solution from the modified field equations and
subsequently analyze its singularity properties through the behavior
of the associated curvature invariants. Section \ref{sec:Thermodynamics-of-the}
is devoted to the thermodynamic analysis of the obtained black hole
solution, where we compute and discuss key thermodynamic quantities
such as the Hawking temperature, entropy, and specific heat. Finally,
in Section \ref{sec:Conclusion}, we summarize our findings and discuss
possible extensions and implications of the results within the context
of modified gravity theories and black hole physics. Throughout this
work, we adopt natural units by setting the gravitational constant
$8\pi G$, the speed of light $c$, the reduced Planck constant $\hbar$,
and the Boltzmann constant $k_{B}$ equal to unity.

\section{\label{sec:Born-Infeld--gravity}Born-Infeld-$f\left(R\right)$ gravity}

Let us briefly review the standard BI-$f\left(R\right)$ theory \citep{makarenko2014born}
(see also \citep{Makarenko:2014nca,odintsov2014born}). In this theory,
the original EiBI gravity theory is combined with an additional $F\left(R\right)$
that depends on the Ricci scalar $R=g^{\mu\nu}R_{\mu\nu}\left(\Gamma\right)$.
To avoid any ghost instabilities, the theory is formulated within
the Palatini formalism (for more detail see \citep{Olmo:2011uz,olmo2012open}),
in which the metric $g_{\mu\nu}$ and the connection $\Gamma_{\beta\gamma}^{\alpha}$
are treated as independent variables. The action for this theory is
given by

\begin{eqnarray}
S & = & \frac{1}{\epsilon}\int\text{d}^{4}x\left[\sqrt{-|g_{\mu\nu}+\epsilon R_{\mu\nu}|}-\lambda\sqrt{-g}\right]+\frac{\alpha}{2}\int\text{d}^{4}x\left[\sqrt{-g}F\left(R\right)\right]+S_{m},\label{eq: action_f(R)}
\end{eqnarray}
where the first term represents the standard BI gravitational Lagrangian
and the second term is an additional function of the Ricci scalar.
$S_{m}$ is the action associated with the matter fields. The parameter
$\epsilon$ is a constant with inverse dimensions to that of a cosmological
constant, and $\lambda$ is a dimensionless constant. Note that $g$
is the determinant of the metric. 

In the limit as $\epsilon\rightarrow0$, the BI Lagrangian reduces
to the standard GR term, and the action simplifies to an $F\left(R\right)$
theory. Alternatively, if we take the limit $\alpha\rightarrow0$,
we retrieve the BI theory. When we take both limits, GR is naturally
recovered. Therefore, the EiBI parameter $\epsilon$ bridges these
different gravity theories. Note that we assume vanishing torsion
and a symmetric Ricci tensor in this formulation.

The field equations can be derived from Eq.(\ref{eq: action_f(R)})
through independent variation concerning the metric and connection.
The variation of this action with respect to the metric tensor leads
to modified metric field equations for the standard BI gravitational
model,
\begin{equation}
\frac{\sqrt{-q}}{\sqrt{-g}}q^{\mu\nu}-\left[\lambda-\frac{\alpha\epsilon}{2}F\left(R\right)\right]g^{\mu\nu}-\alpha\epsilon F_{R}R^{\mu\nu}=-\epsilon T^{\mu\nu},\label{eq: eom_g-1}
\end{equation}
where $F_{R}$ is the derivative of $F\left(R\right)$ with respect
to the Ricci scalar, and $T^{\mu\nu}$ is the standard energy-momentum
tensor. In the above equations, we have used the notation,
\begin{equation}
q_{\mu\nu}=g_{\mu\nu}+\epsilon R_{\mu\nu},\label{eq: def_q}
\end{equation}
and we denoted the inverse of $q_{\mu\nu}$ by $q^{\mu\nu}$ and $q$
represents the determinant of $q_{\mu\nu}$. Similarly, the connection
$\Gamma_{\beta\gamma}^{\alpha}$ variation has the form;

\begin{equation}
\nabla_{\lambda}\left(\sqrt{-q}q^{\mu\nu}+\alpha\sqrt{-g}F_{R}g^{\mu\nu}\right)=0,\label{eq: eom_conn-1}
\end{equation}
where the covariant derivative is taken with respect to the independent
connection, which is defined for a scalar field $\phi$ as
\begin{equation}
\nabla_{\mu}\phi=\partial_{\mu}\phi-\Gamma_{\mu\alpha}^{\alpha}\phi.
\end{equation}

Furthermore, it is well established in the literature on Palatini
$F\left(R\right)$ theories that the independent connection can be
expressed in terms of an auxiliary metric $h_{\mu\nu}$, which is
conformal to $g_{\mu\nu}$ (for more detail, see \citep{Olmo:2011uz}).
Assuming that $q_{\mu\nu}$ is conformally proportional to the metric
tensor as 
\begin{equation}
q_{\mu\nu}=p\left(R\right)g_{\mu\nu},\label{eq: conf_rel}
\end{equation}
where $p\left(R\right)$ is a function of the Ricci scalar, and insert
this expression into (\ref{eq: eom_conn-1}), which yields
\begin{equation}
\nabla_{\lambda}\left[\left(p\left(R\right)+\alpha F_{R}\right)\sqrt{-g}g^{\mu\nu}\right]=0.\label{eq: eom_conn-2}
\end{equation}
We can now define an auxiliary metric tensor 
\begin{equation}
u_{\mu\nu}=\left(p\left(R\right)+\alpha F_{R}\right)g_{\mu\nu},\label{eq: aux_metric}
\end{equation}
such that the above equation reduces to $\nabla_{\beta}\left[\sqrt{-u}u^{\mu\nu}\right]=0$,
here $u^{\mu\nu}$ represents the inverse representation of $u_{\mu\nu}$.
In Einstein\textquoteright s theory the connection equation takes
exactly this form, $\nabla_{\beta}\left[\sqrt{-g}g^{\mu\nu}\right]=0$,
which establishes the compatibility of the connection with the metric,
thus leading to the Levi-Civita connection as a solution (see \citep{Misner:1973prb}
for details) in the torsionless case. Therefore, in our case, we have
\begin{equation}
\Gamma_{\mu\nu}^{\rho}=\frac{1}{2}\left(u^{-1}\right)^{\rho\sigma}\left(u_{\sigma\nu,\mu}+u_{\mu\sigma,\nu}-u_{\mu\nu,\sigma}\right),
\end{equation}
and this provides a complete and exact solution of the connection
equation. 

\section{\label{sec:Black-Hole-Solution}Black Hole Solution}

Black hole solutions play a pivotal role in testing the consistency
and physical implications of gravitational theories, especially in
modified gravity models where nonlinear and higher-curvature corrections
are introduced. In the context of BI-$f\left(R\right)$ gravity, exploring
static and spherically symmetric black hole configurations allows
us to understand how deviations from GR affect space-time near compact
objects. In this section, we construct a black hole solution by starting
from a suitably chosen ansatz and analyzing the field equations. We
begin by assuming a static, spherically symmetric space-time of the
form:

\begin{equation}
\text{d}s^{2}=-f\left(r\right)\text{d}t^{2}+\frac{1}{f\left(r\right)}\text{d}r^{2}+r^{2}\left(\text{d}\theta^{2}+\sin^{2}\theta\text{d}\phi^{2}\right),\label{eq: metric_spherical}
\end{equation}
where $f\left(r\right)$ is the metric function that encapsulates
the gravitational structure of the black hole and reveals the influence
of both the BI nonlinearity and $F\left(R\right)$ gravity corrections. 

Using the conformal relation between the auxiliary metric $q_{\mu\nu}$
and the physical metric $g_{\mu\nu}$ defined in Eq.(\ref{eq: conf_rel})
and Eq.(\ref{eq: def_q}), it follows that the Ricci tensor is necessarily
proportional to the metric tensor, leading to the condition:
\begin{equation}
R_{\mu\nu}\left(\Gamma\right)=\Omega\left(r\right)g_{\mu\nu},\label{eq: Ricci_g}
\end{equation}
where we define $\Omega\left(r\right)=\frac{1}{\epsilon}\left[p\left(R\right)-1\right]$
for simplicity. 

On the other hand, by introducing a redefinition of the metric function
as 
\begin{equation}
u=p\left(R\right)+\alpha F_{R},\label{eq: def_u(r)}
\end{equation}
the auxiliary metric equation Eq.(\ref{eq: aux_metric}) can be considerably
simplified to the form $u_{\mu\nu}=u\left(r\right)g_{\mu\nu}$. Within
this framework, the auxiliary metric becomes conformally related to
the physical metric, and can be explicitly expressed as
\begin{equation}
\text{d}s_{u}^{2}=-u\left(r\right)f\left(r\right)\text{d}t^{2}+\frac{u\left(r\right)}{f\left(r\right)}\text{d}r^{2}+u\left(r\right)r^{2}\left(\text{d}\theta^{2}+\sin^{2}\theta\text{d}\phi^{2}\right).\label{eq: metric_u_mu_nu}
\end{equation}
By utilizing the Ricci tensor components derived from the physical
metric Eq.(\ref{eq: Ricci_g}), the gravitational field equations
yield the following system of coupled nonlinear differential equations:
\begin{equation}
-\frac{f''\left(r\right)}{2}-\frac{f\left(r\right)u''\left(r\right)}{2u\left(r\right)}-\frac{f'\left(r\right)u'\left(r\right)}{u\left(r\right)}-\frac{f'\left(r\right)}{r}-\frac{f\left(r\right)u'\left(r\right)}{ru\left(r\right)}=\Omega\left(R\right),\label{eq: DE_00}
\end{equation}

\begin{equation}
-\frac{f''\left(r\right)}{2}-\frac{3f\left(r\right)u''\left(r\right)}{2u\left(r\right)}-\frac{f'\left(r\right)u'\left(r\right)}{u\left(r\right)}+\frac{3f\left(r\right)u'\left(r\right)^{2}}{2u\left(r\right)^{2}}-\frac{f'\left(r\right)}{r}-\frac{f\left(r\right)u'\left(r\right)}{ru\left(r\right)}=\Omega\left(R\right),\label{eq: DE_11}
\end{equation}

\begin{equation}
-\frac{f\left(r\right)u''\left(r\right)}{2u\left(r\right)}-\frac{f'\left(r\right)u'\left(r\right)}{2u\left(r\right)}-\frac{f'\left(r\right)}{r}-\frac{2f\left(r\right)u'\left(r\right)}{ru\left(r\right)}-\frac{f\left(r\right)}{r^{2}}+\frac{1}{r^{2}}=\Omega\left(R\right).\label{eq: DE_22_33}
\end{equation}
By suitably combining these equations, two independent differential
equations governing the metric functions can be isolated. The first
is a second-order nonlinear differential equation for $u\left(r\right)$

\begin{equation}
\frac{u''}{u}-\frac{3}{2}\left(\frac{u'}{u}\right)^{2}=0,\label{eq: DE_01}
\end{equation}
and the second is;
\begin{equation}
-\frac{f''\left(r\right)}{2}-\frac{f'\left(r\right)u'\left(r\right)}{2u\left(r\right)}+\frac{f\left(r\right)u'\left(r\right)}{ru\left(r\right)}+\frac{f\left(r\right)}{r^{2}}-\frac{1}{r^{2}}=0.\label{eq: DE_03}
\end{equation}
Solving the first equation yields the general form of the auxiliary
function:

\begin{equation}
u\left(r\right)=\frac{4}{\left(C_{1}r+C_{2}\right)^{2}},\label{eq: sol_u(r)}
\end{equation}
where $C_{1}$ and $C_{2}$ are integration constants. In our notation,
we denote the arbitrary integration constants that arise throughout
the analysis as $C_{1},C_{2},...,C_{n}$, with $n\in\mathbb{Z}^{+}$.
Substituting this solution into the second differential equation Eq.(\ref{eq: DE_03}),
the metric function $f\left(r\right)$ can be obtained as:

\begin{equation}
f\left(r\right)=-\frac{\Lambda}{3}r^{2}+\left(\frac{2C_{1}}{C_{2}}+3C_{1}^{2}C_{3}\right)r-\frac{2M}{r}+3C_{1}C_{2}C_{3}+1,\label{eq: metric_function}
\end{equation}
where an effective cosmological constant term is defined as
\begin{equation}
\Lambda=-3C_{4},\label{eq: Lambda_eff}
\end{equation}
that controls the asymptotic (AdS-like) behavior. The inverse radial
term $\propto1/r$ resembles mass effects
\begin{equation}
M=-\frac{C_{2}^{2}C_{3}}{2},\label{eq: Mass_metric_function}
\end{equation}
which implies $C_{3}<0$. The linear term $\propto r$ and constant
terms is a direct result of the extended gravitational dynamics and
encode genuine deviations from general relativity.

In the limit $C_{1}=0$ , for which the function $u\left(r\right)$
(\ref{eq: sol_u(r)}) reduces to a constant, all non-standard contributions
vanish and the metric function (\ref{eq: metric_function}) smoothly
reduces to the Sch--(A)dS solution, confirming the consistency of
the model,

\begin{equation}
f\left(r\right)=-\frac{\Lambda}{3}r^{2}-\frac{2M}{r}+1,\label{eq: metric_function_SdS-1}
\end{equation}
where $\Lambda>0$ corresponds to the de Sitter background, while
$\Lambda<0$ yields the anti--de Sitter geometry. Furthermore, under
this constraint, choosing $C_{2}=2$ leads to the identification of
the auxiliary metric with the space-time metric, $u_{\mu\nu}=g_{\mu\nu}$,
so that the theory reduces to a single-metric description in which
all geometric and physical properties are governed solely by $g_{\mu\nu}$,
as in standard general relativity.

The horizon radius is determined by the condition $f\left(r_{h}\right)=0$
which yields a cubic equation for $r$. Solving this equation reveals
one real root and two complex conjugate roots. The physically relevant
event horizon corresponds to the unique real solution (is also called
outer event horizon), which is expressed in terms of the cosmological
constant $\Lambda$ and the integration constants as:
\begin{equation}
r_{h}=\frac{-C_{2}^{2}\mathcal{A}^{2}+2C_{2}\left(3C_{1}^{2}C_{2}C_{3}+2C_{1}\right)\mathcal{A}-4\left(3\Lambda C_{1}C_{2}^{3}C_{3}+9C_{1}^{4}C_{2}^{2}C_{3}^{2}+\Lambda C_{2}^{2}+12C_{1}^{3}C_{2}C_{3}+4C_{1}^{2}\right)}{2\Lambda C_{2}^{2}\mathcal{A}},\label{eq: Horizon_Radius}
\end{equation}
where the auxiliary function $\mathcal{A}$ is used to write $r_{h}$
in compact form;

\begin{eqnarray}
\mathcal{A} & = & \Biggl\{-\frac{4}{C_{2}^{3}}\biggl(54C_{1}^{6}C_{2}^{3}C_{3}^{3}+27\Lambda C_{1}^{3}C_{2}^{4}C_{3}^{2}+108C_{1}^{5}C_{2}^{2}C_{3}^{2}+3\Lambda^{2}C_{2}^{5}C_{3}+27\Lambda C_{1}^{2}C_{2}^{3}C_{3}+72C_{1}^{4}C_{2}C_{3}+6\Lambda C_{1}C_{2}^{2}\nonumber \\
 &  & +\Lambda C_{2}^{2}\sqrt{(3C_{1}^{3}C_{2}C_{3}+\Lambda C_{2}^{2}+3C_{1}^{2})(27C_{1}^{3}C_{2}^{3}C_{3}^{3}+9\Lambda C_{2}^{4}C_{3}^{2}+27C_{1}^{2}C_{2}^{2}C_{3}^{2}-4)}+16C_{1}^{3}\biggr)\Biggl\}^{1/3}.\label{eq: A_auxiliary_func}
\end{eqnarray}
The existence of real roots for $f(r)=0$ is primarily governed by
the term under the square root within the auxiliary function $\mathcal{A}$
which imposes a well-defined constraint on the integration constants
and the cosmological constant $\Lambda$. Furthermore, by evaluating
the metric function at the event horizon, the mass parameter $M$
can be uniquely expressed in terms of the horizon radius and the integration
constants as:
\begin{equation}
M=-\frac{\Lambda}{6}r_{h}^{3}+\left(\frac{3C_{1}^{2}C_{2}C_{3}+2C_{1}}{2C_{2}}\right)r_{h}^{2}+\left(\frac{1+3C_{1}C_{2}C_{3}}{2}\right)r_{h}.\label{eq: mass_horizon}
\end{equation}

To illustrate the physical viability of the solution, we consider
two representative numerical configurations,
\begin{equation}
\left\{ C_{2}=1,C_{3}=-2\right\} \,\,\,\,\,\,\text{and \ensuremath{\,\,\,\,\,\,\left\{ C_{2}=-1,C_{3}=-2\right\} }},\label{eq: Numerical_Configurations}
\end{equation}
chosen such that the mass parameter is normalized to $M=1$ (\ref{eq: Mass_metric_function})
for convenience. Throughout the analysis, we fix the cosmological
constant to $\Lambda=-0.03$, corresponding to an AdS background,
and use the Sch--AdS case as a reference. The admissible range of
the integration constant $C_{1}$ is determined by requiring the reality
of the square root term appearing in the auxiliary function (\ref{eq: A_auxiliary_func}).
For the configurations above, this requirement yields the following
intervals for which the horizon equation admits a unique real and
positive root:

\begin{equation}
\left\{ -\infty,-0.1853\right\} ,\left\{ -0.0919,0.1138\right\} ,\left\{ 0.4781,\infty\right\} ,\label{eq: C1_limit_1}
\end{equation}
and

\begin{equation}
\left\{ -\infty,-0.4781\right\} ,\left\{ -0.1138,0.0919\right\} ,\left\{ 0.1853,\infty\right\} .\label{eq: C1_limit_2}
\end{equation}
respectively. The numerical evaluation of the central admissible intervals
confirms that the corresponding horizon radius satisfies $r_{h}>0$
consistently resides within the range $1.69\lesssim r_{h}\lesssim2.41$
(in close agreement with the Sch-AdS horizon radius $r_{h}^{Sch}\approx1.93$).
In contrast, the remaining mathematical intervals do not yield physically
meaningful configurations when compared to the Sch--AdS case. To
examine the influence of the $C_{1}$ corrections on the horizon topology,
we focus our analysis on the intermediate regime, specifically 
\begin{equation}
C_{1}\in\left\{ -0.08,-0.03,0.03,0.08\right\} .\label{eq: C1_values}
\end{equation}
Within this range, the curvature singularity at $r=0$ is consistently
cloaked by an event horizon, ensuring a well-defined black hole geometry.

\begin{figure}
\begin{centering}
\includegraphics[width=9cm]{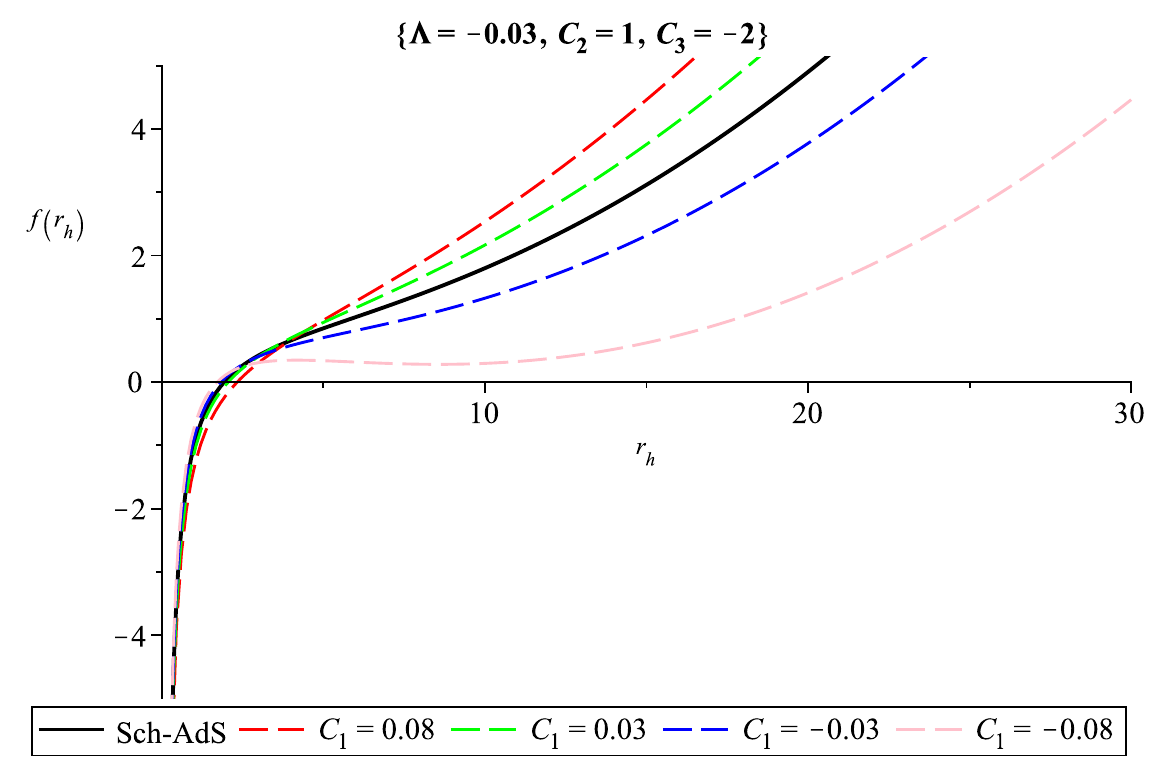}\includegraphics[width=9cm]{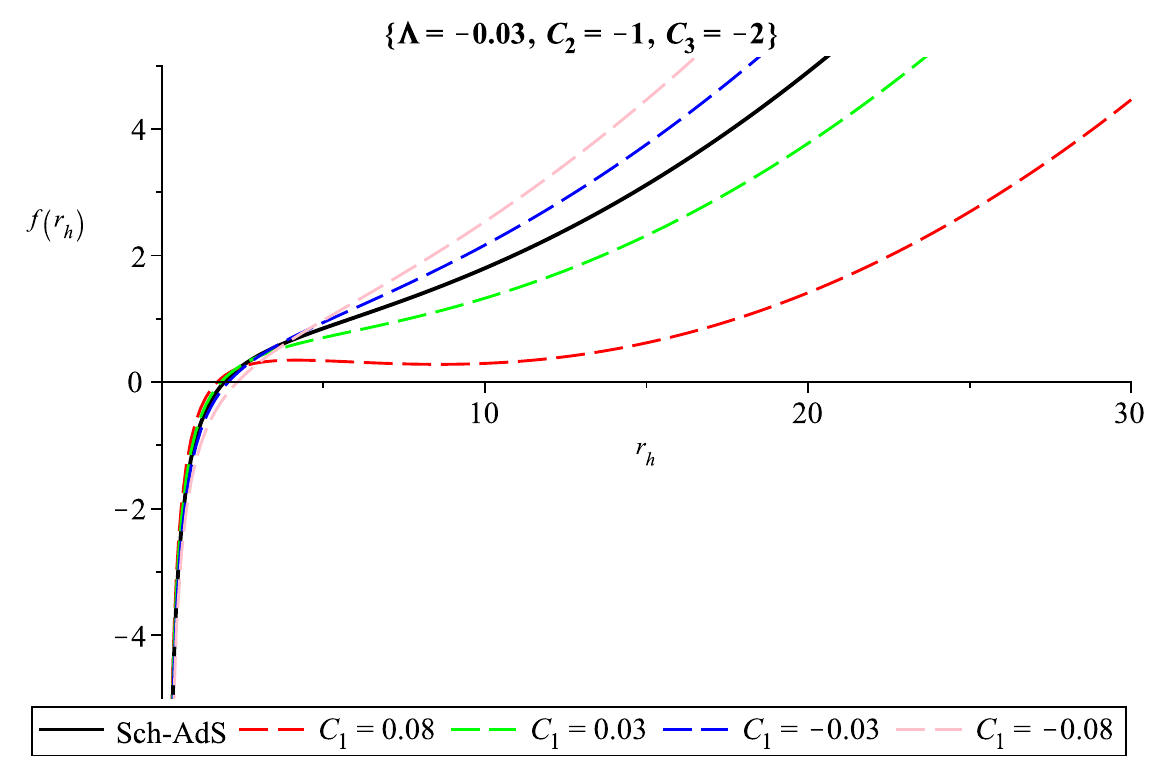}
\par\end{centering}
\caption{(color online) The metric function $f\left(r\right)$ (\ref{eq: metric_function})
is plotted as a function of the horizon radius $r_{H}$ (\ref{eq: Horizon_Radius})
for the set of two configurations in (\ref{eq: Numerical_Configurations}).
The Sch--AdS (black line) solution (\ref{eq: metric_function_SdS-1})
shown for reference. \label{fig: Metric_Function}}
\end{figure}

Figure \ref{fig: Metric_Function} illustrates the behavior of the
metric function $f\left(r\right)$ as a function of the horizon radius
$r_{h}$ for two configurations in (\ref{eq: Numerical_Configurations}).
For the considered parameter space, $f(r)$ admits a unique root at
$r=r_{h}$, identifying a single event horizon. At large scales, $f(r)$
recovers the Sch--AdS asymptotic structure, where the integration
constant $C_{1}$ modulates the horizon radius without introducing
additional roots or geometric pathologies.

Finally, combining the conformal ansatz in (\ref{eq: conf_rel}) with
the definition (\ref{eq: def_q}), one obtains that $p\left(R\right)=1+\frac{\epsilon}{4}R$,
then using the definition of $u\left(r\right)$ (\ref{eq: def_u(r)}),
this relation allows one to straightforwardly obtain
\begin{equation}
F_{R}=-\frac{1}{\alpha}\left(\frac{\epsilon}{4}R-u\left(r\right)+1\right),
\end{equation}
then integrating this expression with respect to $R$, the corresponding
curvature function is derived as
\begin{equation}
F\left(R\right)=-\frac{\epsilon}{8\alpha}R^{2}+\frac{1}{\alpha}\left(u\left(r\right)-1\right)R+C_{5}.\label{eq: F(R)_exact}
\end{equation}

To elucidate the physical content of the theory, we consider the expansion
of the action (\ref{eq: action_f(R)}) in the limit where the coupling
parameter $\epsilon$ is small ($\epsilon\to0$). First, we expand
the determinant inside the square root for the first part in Eq.(\ref{eq: action_f(R)}),
and substitute this back into the first part of the action:
\begin{equation}
\frac{1}{\epsilon}\left[\sqrt{-|g_{\mu\nu}+\epsilon R_{\mu\nu}|}-\lambda\sqrt{-g}\right]\approx\sqrt{-g}\left[\frac{1-\lambda}{\epsilon}+\frac{1}{2}R+\frac{\epsilon}{8}R^{2}-\frac{\epsilon}{4}R_{\mu\nu}R^{\mu\nu}\right].
\end{equation}
Next, we taking into account of the exact solution of $F(R)$, the
action (\ref{eq: action_f(R)}) finally takes the following form,
\begin{equation}
S\approx\int d^{4}x\sqrt{-g}\left[\frac{u(r)}{2}R+\frac{\epsilon}{16}R^{2}-\frac{\epsilon}{4}R_{\mu\nu}R^{\mu\nu}+\Lambda\right]+S_{m}.
\end{equation}
Consequently, the theory reduces to Einstein gravity with an effective
cosmological constant, supplemented by quadratic curvature corrections
characteristic of BI--type modifications. The function $u\left(r\right)$
(\ref{eq: sol_u(r)}) reflects a nontrivial effective gravitational
coupling induced by the modified theory, and the effective cosmological
constant emerges (for the previous definition, see Eq.(\ref{eq: Lambda_eff}))
as a combination of the model parameters ($\epsilon$, $\alpha$ and
$\lambda$) and integration constant $C_{5}$, 

\begin{equation}
\Lambda=\frac{1-\lambda}{\epsilon}+\frac{\alpha C_{5}}{2}.
\end{equation}
ensuring consistency with the exact solution. 

\subsection{Singularity Analysis from Curvature Invariants}

To determine the regularity of the space-time described by the metric
function in (\ref{eq: metric_function}), we examine curvature scalars
constructed from the Riemann tensor. The presence of a physical curvature
singularity is characterized by the divergence of the Ricci scalar
$R$, the Ricci tensor squared $R_{\mu\nu}R^{\mu\nu}$, and the Kretschmann
scalar $\mathcal{K}=R_{\mu\nu\rho\sigma}R^{\mu\nu\rho\sigma}$. For
the present solution, these invariants take the form;
\begin{equation}
R=4\Lambda-\frac{18C_{1}^{2}C_{3}-12C_{1}C_{2}^{-1}}{r}-\frac{6C_{2}C_{1}C_{3}}{r^{2}},
\end{equation}

\begin{eqnarray}
R_{\mu\nu}R^{\mu\nu} & = & 4\Lambda^{2}-\frac{8\Lambda\left(4C_{1}^{2}C_{3}+3C_{1}C_{2}^{-1}\right)}{r}+\frac{120C_{1}^{3}C_{2}^{-1}C_{3}+90C_{1}^{4}C_{3}^{2}+40C_{1}^{2}C_{2}^{-2}-12C_{2}C_{1}C_{3}\Lambda}{r^{2}}\nonumber \\
 &  & +\frac{72C_{2}C_{1}^{3}C_{3}^{2}+48C_{1}^{2}C_{3}}{r^{3}}+\frac{18C_{2}^{2}C_{1}^{2}C_{3}^{2}}{r^{4}},
\end{eqnarray}
and

\begin{eqnarray}
\mathcal{K} & = & \frac{8\Lambda^{2}}{3}-\frac{8\Lambda\left(3C_{1}^{3}C_{3}+2C_{1}C_{2}^{-1}\right)}{r}+\frac{96C_{1}^{3}C_{2}^{-1}C_{3}+72C_{1}^{4}C_{3}^{2}+32C_{1}^{2}C_{2}^{-2}-8C_{1}C_{2}C_{3}\Lambda}{r^{2}}\nonumber \\
 &  & +\frac{72C_{2}C_{1}^{3}C_{3}^{2}+48C_{1}^{2}C_{3}}{r^{3}}+\frac{36C_{2}^{2}C_{1}^{2}C_{3}^{2}}{r^{4}}+\frac{24C_{2}^{3}C_{1}C_{3}^{2}}{r^{5}}+\frac{12C_{2}^{4}C_{3}^{2}}{r^{6}}.
\end{eqnarray}
Now, let us analyze the asymptotic behavior of these quantities in
the limit $r\rightarrow0$. The Ricci scalar diverges as $R\sim r^{-2}$
, while the Ricci tensor squared contains terms diverging up to $r^{-4}$.
Most notably, the Kretschmann scalar exhibits the strongest divergence,

\begin{equation}
\mathcal{K}\sim\frac{12C_{2}^{4}C_{3}^{2}}{r^{6}},\,\,\,\,\,\,\,r\rightarrow0\label{eq: Kretschmann_r_0}
\end{equation}
which is of the same leading order as the Sch-AdS cases. This shows
that the central singularity is of Schwarzschild type, characterized
by a strong $r^{-6}$ divergence. Subleading divergences reflect the
modified gravitational structure but do not alter the leading singular
behavior. Conversely, at large radii $r\to\infty$, all terms depending
on the integration constants $C_{i}$ decay, and the invariants approach;
$R\to4\Lambda$, $R_{\mu\nu}R^{\mu\nu}\to4\Lambda^{2}$, and $\mathcal{K}\to\frac{8}{3}\Lambda^{2}$
confirming that the space-time is asymptotically AdS. 

\section{\label{sec:Thermodynamics-of-the}Thermodynamics of the model}

\begin{figure}
\begin{centering}
\includegraphics[width=9cm]{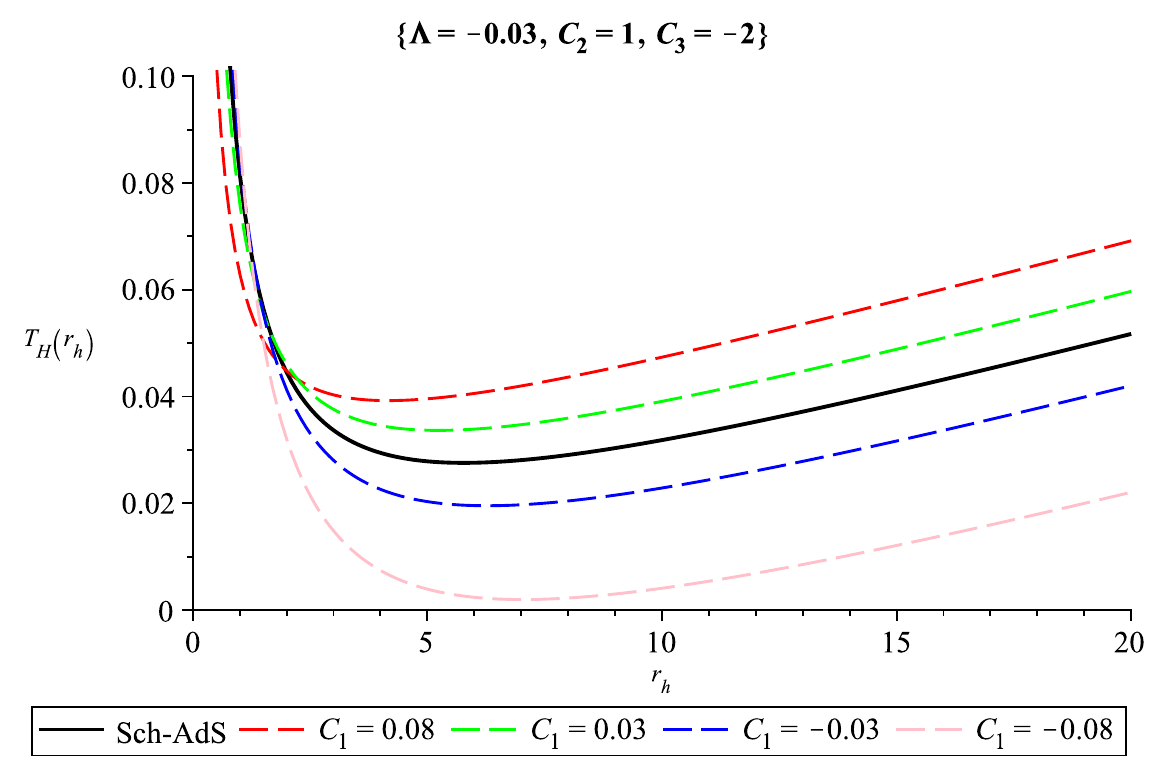}\includegraphics[width=9cm]{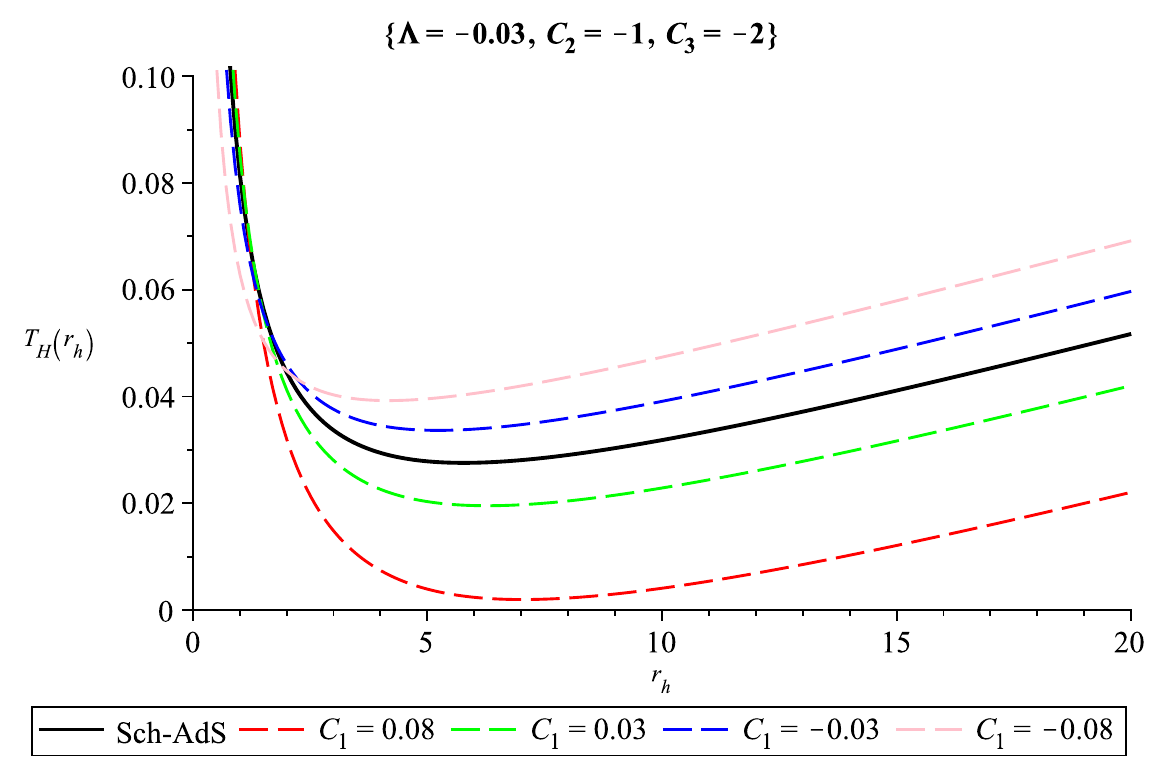}
\par\end{centering}
\caption{(color online) The Hawking temperature (\ref{eq: metric_function})
is plotted as a function of the horizon radius $r_{H}$ (\ref{eq: Horizon_Radius})
for the parameter sets in (\ref{eq: Numerical_Configurations}). The
Sch--AdS (black line) solution (\ref{eq: metric_function_SdS-1})
shown for a reference. \label{fig: Hawking_Temperature}}
\end{figure}

In this section, we investigate the thermodynamic properties of the
black hole solution derived from the BI-$f\left(R\right)$ gravitational
framework. Thermodynamic analysis provides valuable insights into
the interplay between gravity, quantum field theory, and the fundamental
laws of black hole mechanics. We begin our thermodynamical analysis
by computing the surface gravity, $\kappa$, which quantifies the
gravitational acceleration at the event horizon and serves as a key
quantity in the determination of black hole temperature. For a static
and spherically symmetric space-time, the surface gravity is defined
as \citep{Visser:1992Dirty}:
\begin{eqnarray}
\kappa & = & \lim_{r\rightarrow r_{H}}\left(\frac{1}{2}\frac{1}{\sqrt{g_{tt}g_{rr}}}\frac{\partial}{\partial r}g_{tt}\right).
\end{eqnarray}
In the context of the metric given in (\ref{eq: metric_spherical}),
and taking into account of the mass expression (\ref{eq: mass_horizon}),
this expression simplifies to:
\begin{eqnarray}
\kappa & = & \lim_{r\rightarrow r_{H}}\left(\frac{1}{2}\frac{\partial}{\partial r}g_{tt}\right)=-\frac{\Lambda r_{H}}{2}+\frac{\left(3C_{1}C_{2}C_{3}+1\right)}{2r_{H}}+C_{1}\left(\frac{2}{C_{2}}+3C_{1}C_{3}\right).
\end{eqnarray}
For the parameter values considered in (\ref{eq: Numerical_Configurations})
and (\ref{eq: C1_values}), the surface gravity evaluated at the event
horizon is strictly positive, ensuring that the Hawking temperature
$T_{H}$ is well defined. The Hawking temperature, which characterizes
the thermal radiation emitted by the black hole as predicted by quantum
field theory in curved space-time \citep{Hawking:1975vcx}, is proportional
to the surface gravity $\kappa$ evaluated at the outer horizon $r=r_{h}$
and is given by
\begin{equation}
T_{H}=\frac{\kappa}{2\pi}=\frac{\hbar}{4\pi k_{B}}\left[-\Lambda r_{H}+\frac{\left(3C_{1}C_{2}C_{3}+1\right)}{r_{H}}+2C_{1}\left(\frac{2}{C_{2}}+3C_{1}C_{3}\right)\right].\label{eq: temp_hawk}
\end{equation}
Figure \ref{fig: Hawking_Temperature} illustrates the thermodynamic
behavior of black holes in our modified gravity theory, showing the
Hawking temperature ($T_{H}$) (\ref{eq: temp_hawk}) versus the horizon
radius ($r_{H}$) for the parameter sets considered in this work (\ref{eq: Numerical_Configurations}).
All solutions display the characteristic Hawking-Page transition \citep{Hawking:1983ThermodynamicsAdS},
where $T_{H}$ reaches a minimum, separating unstable (small $r_{h}$)
from stable (large $r_{h}$) black holes.

The entropy associated with the black hole horizon can be determined
from the first law of black hole thermodynamics. For a static configuration,
the first law reduces to 
\begin{eqnarray}
dM & = & T_{H}dS,
\end{eqnarray}
where $S$ denotes the entropy associated with the horizon. Considering
the mass $M$ (\ref{eq: mass_horizon}) and the Hawking temperature
$T_{H}$ (\ref{eq: temp_hawk}) as functions of the horizon radius
$r_{h}$, one finds

\begin{equation}
dS=\frac{1}{T_{H}}\frac{dM}{dr_{h}}dr_{h}.
\end{equation}
Integrating with respect to $r_{h}$, the entropy is obtained as
\begin{equation}
S\left(r_{h}\right)=\pi r_{h}^{2}.\label{eq: Entropy}
\end{equation}
In Palatini $F\left(R\right)$ theories of gravity, the black hole
entropy is generally expected to receive a correction proportional
to the derivative $F_{R}=\frac{dF\left(R\right)}{dR}$, leading to
the generalized relation $S=\frac{A}{4}F_{R}$ \citep{Brevik:2004Entropy,Cognola:2005Oneloop},
and, in the presence of BI--type extensions, to an even more complex
Wald entropy functional \citep{Ozen:2017Entropy}. In contrast, the
entropy obtained from the thermodynamic integration in the present
BI--$F\left(R\right)$ framework obeys the standard Bekenstein--Hawking
form \citep{Bekenstein1973,Gibbons:1977Cosmological} $S=A/4$ (where
the horizon area is $A=4\pi r_{h}^{2}$), with no explicit contribution
arising from neither the BI sector nor the $F\left(R\right)$ modification.
This result can be understood as a consequence of the specific structure
of the BI--$F\left(R\right)$ theory in the Palatini formulation,
where the scalar curvature is determined algebraically by the field
equations and does not represent an independent dynamical degree of
freedom at the horizon.

To assess local thermodynamic stability, we compute the specific heat
$C$ at constant horizon radius:

\begin{equation}
C=T\left(\frac{\partial S}{\partial T}\right)_{T=T_{H}}.
\end{equation}
Substituting the expressions for $S$ and $T_{H}$, we obtain:
\begin{eqnarray}
C & = & -2\pi r_{h}^{2}\left(\frac{2C_{1}\left(3C_{1}^{2}C_{2}C_{3}+2\right)r_{h}+C_{2}\left(-\Lambda r_{h}^{2}+3C_{1}C_{2}C_{3}+1\right)}{C_{2}\left(\Lambda r_{h}^{2}+3C_{1}C_{2}C_{3}+1\right)}\right).\label{eq: Heat_Capacity}
\end{eqnarray}
The sign of the specific heat $C$ is a diagnostic tool for thermodynamic
stability: a positive value indicates local stability, while a negative
value suggests instability and possible phase transitions. 

\begin{figure}
\begin{centering}
\includegraphics[width=9cm]{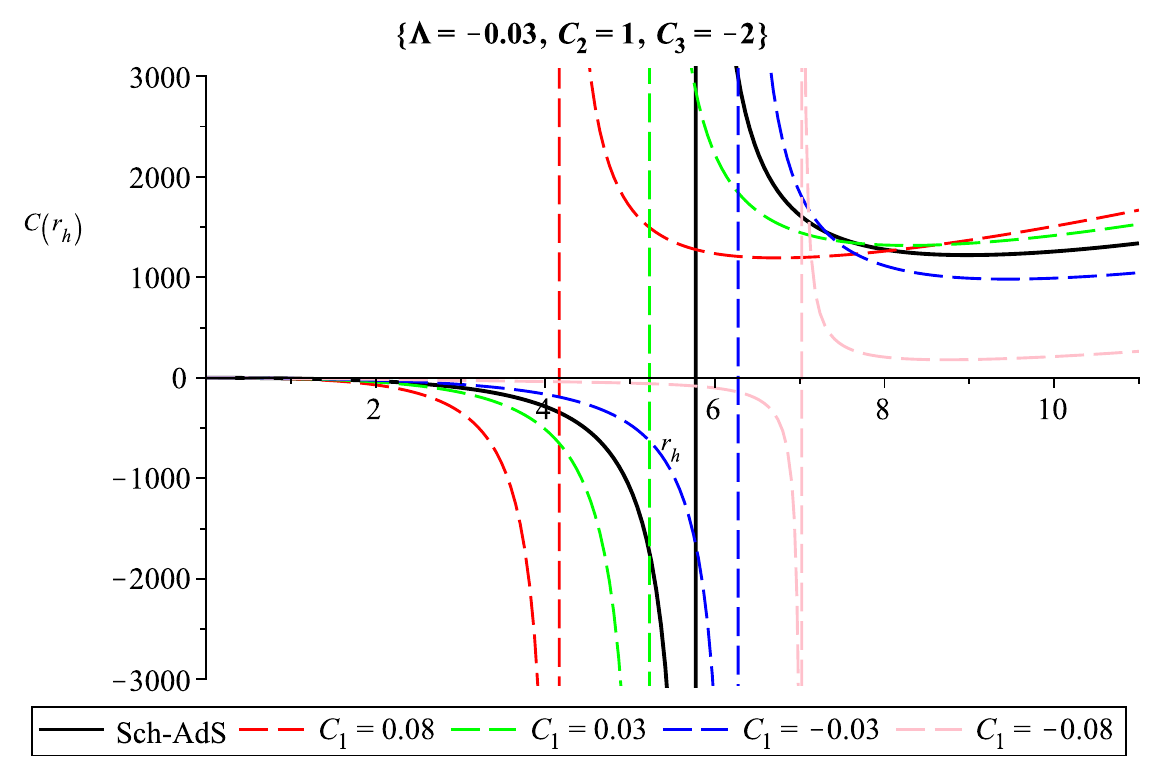}\includegraphics[width=9cm]{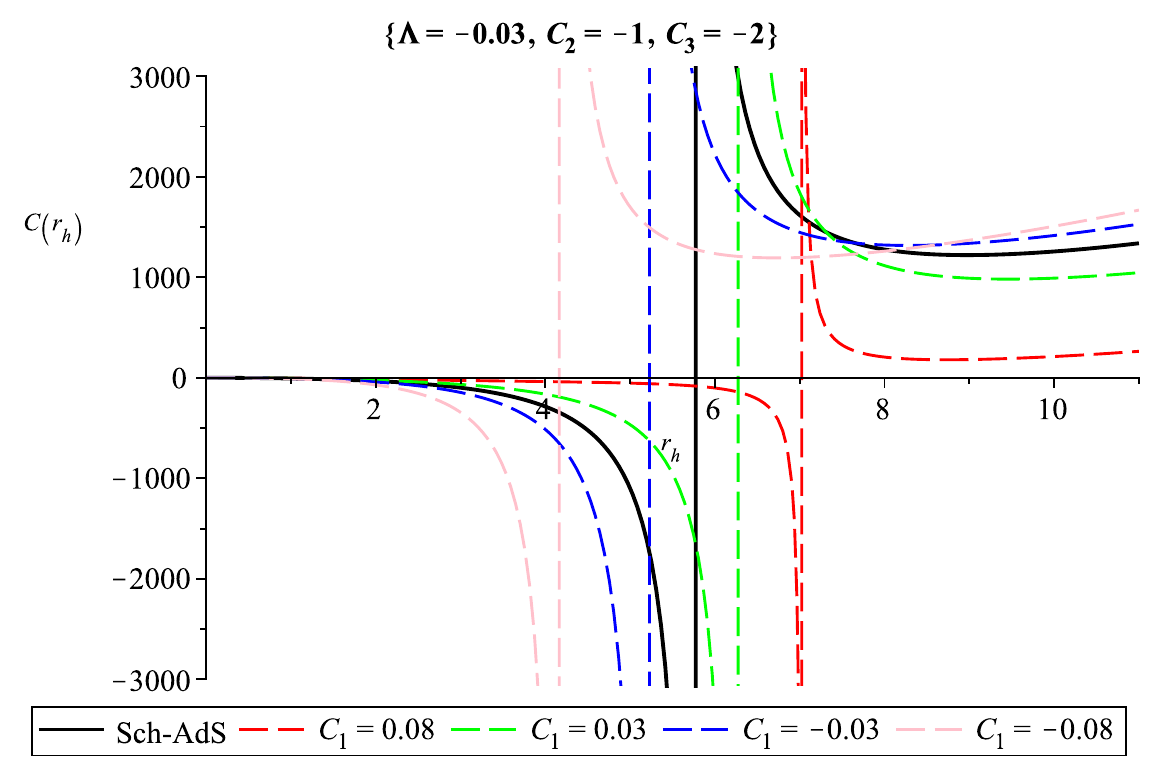}
\par\end{centering}
\caption{(color online) The heat capacity $C$ (\ref{eq: temp_hawk}) as a
function of the horizon radius $r_{H}$ under the considered parameter
sets in (\ref{eq: Numerical_Configurations}). The plots include comparisons
with the Sch-AdS (black solid line) black hole. The vertical divergences
in $C$ signal phase transitions, while the sign of $C$ indicates
thermodynamic stability. \label{fig: Heat_Capacity}}
\end{figure}

In Figure \ref{fig: Heat_Capacity}, we present the variation of the
specific heat capacity as a function of $r_{H}$, highlighting the
conditions under which the black hole becomes thermodynamically stable
or unstable. The analysis reveals that the stability behavior is strongly
dependent on the integration constants. For instance, variations in
the integration constant $C_{1}$ shift the location of these divergences
and alter the size of the stable branches, showing that the model
parameters significantly influence the stability structure of the
black holes.

Moreover, divergences in the heat capacity $C$ are associated with
second-order phase transitions, which are often linked to critical
points in black hole thermodynamics. Such divergences arise when the
denominator of Eq.(\ref{eq: Heat_Capacity}) vanishes, leading to
the analytic condition,

\begin{equation}
\Lambda r_{c}^{2}+3C_{1}C_{2}C_{3}+1=0,
\end{equation}
which defines the critical horizon radius,

\begin{equation}
r_{c}=\sqrt{-\frac{3C_{1}C_{2}C_{3}+1}{\Lambda}}.
\end{equation}
In the AdS regime ($\Lambda<0$), the critical radius is real provided
$3C_{1}C_{2}C_{3}+1>0$, a condition satisfied by the parameter sets
considered here. Evaluating the Hawking temperature at $r_{h}=r_{c}$
then yields the corresponding critical temperature, which characterizes
the onset of the thermodynamic phase transition,

\begin{equation}
T_{H}\left(r_{c}\right)=\frac{\hbar}{2\pi k_{B}}\left[3C_{1}^{2}C_{3}+\frac{2C_{1}}{C_{2}}+\frac{\left(3C_{1}C_{2}C_{3}+1\right)}{\sqrt{-\frac{3C_{1}C_{2}C_{3}+1}{\Lambda}}}\right].\label{eq: Th_critical}
\end{equation}
Requiring the critical temperature to be non-negative, $T_{H}\left(r_{c}\right)\geq0$,
imposes physical bounds on the integration constant $C_{1}$. Solving
(\ref{eq: Th_critical}) with respect to $C_{1}$, for the two configurations
considered in (\ref{eq: Numerical_Configurations}), the allowed numerical
ranges for $C_{1}$ are found to be $C_{1}>-0.0848$ and $C_{1}<0.0848$,
respectively, in accordance with Eqs.(\ref{eq: C1_limit_1}) and (\ref{eq: C1_limit_2}).
These bounds define the thermodynamically consistent parameter space
for which the solution describes a physically viable black hole with
a well-defined thermal state. For instance, adopting the first configuration
with $C_{1}=0.08$, we obtain $r_{c}\approx4.16$, in excellent agreement
with the minima of the Hawking temperature shown in Fig. (\ref{fig: Hawking_Temperature}),
thereby confirming that the divergence of the heat capacity coincides
with the transition from negative to positive specific heat.

\section{\label{sec:Conclusion}Conclusion}

In this work, we have investigated static, spherically symmetric black
hole solutions in the framework of BI--$f\left(R\right)$ gravity,
a class of modified theories motivated by BI--type nonlinearities
and extensions of general relativity. Our analysis focused on constructing
an exact black hole geometry and examining its geometric and thermodynamic
properties within an asymptotically AdS background.

The resulting metric function (\ref{eq: metric_function}) preserves
the asymptotically AdS structure of the space-time, as illustrated
in Fig. \ref{fig: Metric_Function}. The integration constant $C_{1}$
parametrizes deviations from the Sch--AdS geometry, inducing quantitative
shifts in the horizon structure while smoothly recovering the Sch--AdS
limit as $C_{1}\rightarrow0$. Although BI--type gravity theories
are often associated with singularity regularization, the black hole
solutions obtained in this work do not eliminate the central curvature
singularity. In particular, the Kretschmann scalar diverges as $r^{-6}$
(\ref{eq: Kretschmann_r_0}), indicating a Sch-type curvature singularity
and geodesic incompleteness. This result highlights that the regularizing
properties of BI--like modifications are not universal and depend
sensitively on the underlying gravitational dynamics.

The Hawking temperature (\ref{eq: temp_hawk}) remains positive and
exhibits a minimum as a function of the horizon radius, separating
small and large black hole branches in close analogy with the Sch--AdS
case (see Fig. (\ref{fig: Hawking_Temperature})). This behavior signals
the onset of thermodynamic stability for sufficiently large black
holes. The heat capacity (\ref{eq: Heat_Capacity}) further supports
this interpretation: its sign change distinguishes unstable and stable
phases, while its divergence marks second-order phase transitions
(see Fig. (\ref{fig: Heat_Capacity})). Variations of $C_{1}$ shift
the critical points but preserve the overall phase structure, confirming
that the modified gravity effects introduce controlled deviations
without altering the qualitative thermodynamic behavior of the black
hole solutions. 

In summary, BI-$f\left(R\right)$ gravity admits asymptotically AdS
black hole solutions with well-defined thermodynamic behavior closely
resembling that of Sch--AdS black holes, while exhibiting distinctive
quantitative modifications governed by the integration constants.
These results provide a consistent starting point for further studies,
including extensions to charged or rotating configurations, higher-dimensional
generalizations, and potential implications for modified gravity phenomenology.

\section*{Declaration of competing interest }

The authors declare that they have no known competing financial interests
or personal relationships that could have appeared to influence the
work reported in this paper.

\section*{Declaration of generative AI and AI-assisted technologies in the
writing process}

During the preparation of this work the author used ChatGPT and Gemini
in order to improve readability of the text. After using this tool/service,
the author reviewed and edited the content as needed and takes full
responsibility for the content of the publication.

\section*{Data availability }

No data was used for the research described in the article.

\bibliography{BIFR_BH_R1}

@article{makarenko2014born,
  title = "{Born-Infeld-$f(R)$ gravity}",
  author = {Makarenko, Andrey N. and Odintsov, Sergei D. and Olmo, Gonzalo J.},
  journal = {Phys. Rev. D},
  volume = {90},
  issue = {2},
  pages = {024066},
  numpages = {15},
  year = {2014},
  month = {Jul},
  publisher = {American Physical Society},
  doi = {10.1103/PhysRevD.90.024066},
  url = {https://link.aps.org/doi/10.1103/PhysRevD.90.024066}
}

@article{Makarenko:2014nca,
    author = "Makarenko, Andrey N. and Odintsov, Sergei D. and Olmo, Gonzalo J.",
    title = "{Little Rip, $\Lambda$CDM and singular dark energy cosmology from Born-Infeld-$f(R)$ gravity}",
    reportNumber = "IFIC-14-27",
    doi = "10.1016/j.physletb.2014.05.024",
    journal = "Phys. Lett. B",
    volume = "734",
    pages = "36--40",
    year = "2014"
}

@article{Elizalde:2016vsd,
    author = "Elizalde, Emilio and Makarenko, Andrey N.",
    title = "{Singular inflation from Born\textendash{}Infeld-f (R) gravity}",
    doi = "10.1142/S0217732316501492",
    journal = "Mod. Phys. Lett. A",
    volume = "31",
    number = "24",
    pages = "1650149",
    year = "2016"
}

@article{Banados2010eddington,
  title = {Eddington's Theory of Gravity and Its Progeny},
  author = {Ba\~nados, M\'aximo and Ferreira, Pedro G.},
    doi = "10.1103/PhysRevLett.105.011101",
    journal = "Phys. Rev. Lett.",
    volume = "105",
    pages = "011101",
    year = "2010",
    note = "[Erratum: Phys.Rev.Lett. 113, 119901 (2014)]"
 }

@article{born1934foundations,
    author = "Born, M. and Infeld, L.",
    title = "{Foundations of the new field theory}",
    doi = "10.1098/rspa.1934.0059",
    journal = "Proc. Roy. Soc. Lond. A",
    volume = "144",
    number = "852",
    pages = "425--451",
    year = "1934"
}

@article{vollick2004palatini,
    author = "Vollick, Dan N.",
    title = "{Palatini approach to Born-Infeld-Einstein theory and a geometric description of electrodynamics}",
    doi = "10.1103/PhysRevD.69.064030",
    journal = "Phys. Rev. D",
    volume = "69",
    pages = "064030",
    year = "2004"
}

@article{vollick2005born,
    author = "Vollick, Dan N.",
    title = "{Born-Infeld-Einstein theory with matter}",
    doi = "10.1103/PhysRevD.72.084026",
    journal = "Phys. Rev. D",
    volume = "72",
    pages = "084026",
    year = "2005"
}

@article{deser1998born,
    author = "Deser, Stanley and Gibbons, G. W.",
    title = "{Born-Infeld-Einstein actions?}",
    reportNumber = "BRX-TH-430",
    doi = "10.1088/0264-9381/15/5/001",
    journal = "Class. Quant. Grav.",
    volume = "15",
    pages = "L35--L39",
    year = "1998"
}

@book{eddington1923mathematical,
  title={The Mathematical Theory of Relativity},
  author={Eddington, Arthur Stanley},
  year={1923},
  publisher={The Cambridge University Press}
}

@article{odintsov2014born,
    author = "Odintsov, Sergei D. and Olmo, Gonzalo J. and Rubiera-Garcia, D.",
    title = "{Born-Infeld gravity and its functional extensions}",
    reportNumber = "IFIC-14-50",
    doi = "10.1103/PhysRevD.90.044003",
    journal = "Phys. Rev. D",
    volume = "90",
    pages = "044003",
    year = "2014"
}

@article{nojiri2007introduction,
  title={Introduction to modified gravity and gravitational alternative for dark energy},
  author={Nojiri, Shin'ichi and Odintsov, Sergei D},
  journal={Int. J. Geom. Methods Mod. Phys.},
  volume={4},
  number={01},
  pages={115--145},
  year={2007},
  doi = "10.1142/S0219887807001928",
  publisher={World Scientific}
}

@article{nojiri2011unified,
    author = "Nojiri, Shin'ichi and Odintsov, Sergei D.",
    title = "{Unified cosmic history in modified gravity: from F(R) theory to Lorentz non-invariant models}",
    doi = "10.1016/j.physrep.2011.04.001",
    journal = "Phys. Rept.",
    volume = "505",
    pages = "59--144",
    year = "2011"
}

@article{Bekenstein1973,
  title = {Black Holes and Entropy},
  author = {Bekenstein, Jacob D.},
  journal = {Phys. Rev. D},
  volume = {7},
  issue = {8},
  pages = {2333--2346},
  numpages = {0},
  year = {1973},
  month = {Apr},
  publisher = {American Physical Society},
  doi = {10.1103/PhysRevD.7.2333},
  url = {https://link.aps.org/doi/10.1103/PhysRevD.7.2333}
}

@article{Hawking:1975vcx,
    author = "Hawking, S. W.",
    editor = "Gibbons, G. W. and Hawking, S. W.",
    title = "{Particle Creation by Black Holes}",
    doi = "10.1007/BF02345020",
    journal = "Commun. Math. Phys.",
    volume = "43",
    pages = "199--220",
    year = "1975",
    note = "[Erratum: Commun.Math.Phys. 46, 206 (1976)]"
}

@article{Olmo:2011uz,
    author = "Olmo, Gonzalo J.",
    title = "{Palatini Approach to Modified Gravity: f(R) Theories and Beyond}",
    doi = "10.1142/S0218271811018925",
    journal = "Int. J. Mod. Phys. D",
    volume = "20",
    pages = "413--462",
    year = "2011"
}

@book{olmo2012open,
    editor = "Olmo, Gonzalo J. and Olmo, Gonzalo J.",
    title = "{Open Questions in Cosmology}",
    doi = "10.5772/45746",
    isbn = "978-953-51-0880-1",
    publisher = "InTech",
    month = "11",
    year = "2012"
}

@article{jimenez2018born,
    author = "Beltran Jimenez, Jose and Heisenberg, Lavinia and Olmo, Gonzalo J. and Rubiera-Garcia, Diego",
    title = "{Born\textendash{}Infeld inspired modifications of gravity}",
    reportNumber = "IFIC-17-23",
    doi = "10.1016/j.physrep.2017.11.001",
    journal = "Phys. Rept.",
    volume = "727",
    pages = "1--129",
    year = "2018"
}

@book{Misner:1973prb,
    author = "Misner, Charles W. and Thorne, K. S. and Wheeler, J. A.",
    title = "{Gravitation}",
    isbn = "978-0-7167-0344-0, 978-0-691-17779-3",
    publisher = "W. H. Freeman",
    address = "San Francisco",
    year = "1973"
}

@article{Nojiri:2017ncd,
    author = "Nojiri, S. and Odintsov, S. D. and Oikonomou, V. K.",
    title = "{Modified Gravity Theories on a Nutshell: Inflation, Bounce and Late-time Evolution}",
    reportNumber = "PHYS.REPT.-692-(2017)-1-104, Phys.Rept. 692 (2017) 1-104",
    doi = "10.1016/j.physrep.2017.06.001",
    journal = "Phys. Rept.",
    volume = "692",
    pages = "1--104",
    year = "2017"
}

@article{Avelino:2012BIcosmologies,
    author = "Avelino, P. P. and Ferreira, R. Z.",
    title = "{Bouncing Eddington-inspired Born-Infeld cosmologies: an alternative to Inflation ?}",
    doi = "10.1103/PhysRevD.86.041501",
    journal = "Phys. Rev. D",
    volume = "86",
    pages = "041501",
    year = "2012"
}

@article{Kibaroglu:2024UnimodularBI,
    author = "Kibaro\u{g}lu, Salih and Odintsov, Sergei D. and Paul, Tanmoy",
    title = "{Cosmology of unimodular Born\textendash{}Infeld-$f(R)$ gravity}",
    doi = "10.1016/j.dark.2024.101445",
    journal = "Phys. Dark Univ.",
    volume = "44",
    pages = "101445",
    year = "2024"
}

@article{Kibaroglu:2023BIdS,
    author = "Kibaro\u{g}lu, Salih",
    title = "{Born\textendash{}Infeld-$f(R)$ gravity with de Sitter solutions}",
    doi = "10.1142/S0219887823501414",
    journal = "Int. J. Geom. Meth. Mod. Phys.",
    volume = "20",
    number = "08",
    pages = "2350141",
    year = "2023"
}

@article{Kibaroglu:2023CosmBI,
    author = "Kibaro\u{g}lu, Salih and Elizalde, Emilio",
    title = "{Cosmological implications of Born\textendash{}Infeld-$f(R)$ gravity}",
    doi = "10.1142/S0218271823500128",
    journal = "Int. J. Mod. Phys. D",
    volume = "32",
    number = "03",
    pages = "2350012",
    year = "2023"
}

@article{Du:2014Large,
    author = "Du, Xiao-Long and Yang, Ke and Meng, Xin-He and Liu, Yu-Xiao",
    title = "{Large Scale Structure Formation in Eddington-inspired Born-Infeld Gravity}",
    doi = "10.1103/PhysRevD.90.044054",
    journal = "Phys. Rev. D",
    volume = "90",
    pages = "044054",
    year = "2014"
}

@article{Kim:2014Origin,
    author = "Kim, Hyeong-Chan",
    title = "{Origin of the universe: A hint from Eddington-inspired Born-Infeld gravity}",
    doi = "10.3938/jkps.65.840",
    journal = "J. Korean Phys. Soc.",
    volume = "65",
    number = "6",
    pages = "840--845",
    year = "2014"
}

@article{Kruglov:2013Modified,
    author = "Kruglov, S. I.",
    title = "{Modified arctan-gravity model mimicking a cosmological constant}",
    doi = "10.1103/PhysRevD.89.064004",
    journal = "Phys. Rev. D",
    volume = "89",
    number = "6",
    pages = "064004",
    year = "2014"
}

@article{Yang:2013Linear,
    author = "Yang, Ke and Du, Xiao-Long and Liu, Yu-Xiao",
    title = "{Linear perturbations in Eddington-inspired Born-Infeld gravity}",
    doi = "10.1103/PhysRevD.88.124037",
    journal = "Phys. Rev. D",
    volume = "88",
    pages = "124037",
    year = "2013"
}

@article{Escamilla-Rivera:2012Tensor,
    author = "Escamilla-Rivera, Celia and Banados, Maximo and Ferreira, Pedro G.",
    title = "{A tensor instability in the Eddington inspired Born-Infeld Theory of Gravity}",
    doi = "10.1103/PhysRevD.85.087302",
    journal = "Phys. Rev. D",
    volume = "85",
    pages = "087302",
    year = "2012"
}

@article{Cho:2012Universe,
    author = "Cho, Inyong and Kim, Hyeong-Chan and Moon, Taeyoon",
    title = "{Universe Driven by Perfect Fluid in Eddington-inspired Born-Infeld Gravity}",
    doi = "10.1103/PhysRevD.86.084018",
    journal = "Phys. Rev. D",
    volume = "86",
    pages = "084018",
    year = "2012"
}

@article{Scargill:2012Cosmology,
    author = "Scargill, James H. C. and Banados, Maximo and Ferreira, Pedro G.",
    title = "{Cosmology with Eddington-inspired Gravity}",
    doi = "10.1103/PhysRevD.86.103533",
    journal = "Phys. Rev. D",
    volume = "86",
    pages = "103533",
    year = "2012"
}

@article{Olmo:2014GeonicBH,
    author = "Olmo, Gonzalo J. and Rubiera-Garcia, D. and Sanchis-Alepuz, Helios",
    title = "{Geonic black holes and remnants in Eddington-inspired Born-Infeld gravity}",
    doi = "10.1140/epjc/s10052-014-2804-8",
    journal = "Eur. Phys. J. C",
    volume = "74",
    pages = "2804",
    year = "2014"
}

@article{Lobo:2014Microscopic,
    author = "Lobo, Francisco S. N. and Olmo, Gonzalo J. and Rubiera-Garcia, D.",
    title = "{Microscopic wormholes and the geometry of entanglement}",
    doi = "10.1140/epjc/s10052-014-2924-1",
    journal = "Eur. Phys. J. C",
    volume = "74",
    number = "6",
    pages = "2924",
    year = "2014"
}

@article{Harko:2015Wormhole,
    author = "Harko, Tiberiu and Lobo, Francisco S. N. and Mak, M. K. and Sushkov, Sergey V.",
    title = "{Wormhole geometries in Eddington-Inspired Born\textendash{}Infeld gravity}",
    doi = "10.1142/S0217732315501904",
    journal = "Mod. Phys. Lett. A",
    volume = "30",
    number = "35",
    pages = "1550190",
    year = "2015"
}

@article{Gibbons:1977Cosmological,
    author = "Gibbons, G. W. and Hawking, S. W.",
    title = "{Cosmological Event Horizons, Thermodynamics, and Particle Creation}",
    doi = "10.1103/PhysRevD.15.2738",
    journal = "Phys. Rev. D",
    volume = "15",
    pages = "2738--2751",
    year = "1977"
}

@article{Visser:1992Dirty,
    author = "Visser, Matt",
    title = "{Dirty black holes: Thermodynamics and horizon structure}",
    doi = "10.1103/PhysRevD.46.2445",
    journal = "Phys. Rev. D",
    volume = "46",
    pages = "2445--2451",
    year = "1992"
}

@article{Kibaroglu:2025AnisotropicBI,
    author = "Kibaro{\u{g}}lu, Salih",
    title = "{Anisotropic Born{\textendash}Infeld-f(R) cosmologies}",
    doi = "10.1016/j.dark.2024.101784",
    journal = "Phys. Dark Univ.",
    volume = "47",
    pages = "101784",
    year = "2025"
}

@article{Avelino:2016InnerStructure,
    author = "Avelino, P. P.",
    title = "{Inner Structure of Black Holes in Eddington-inspired Born-Infeld gravity: the role of mass inflation}",
    doi = "10.1103/PhysRevD.93.044067",
    journal = "Phys. Rev. D",
    volume = "93",
    number = "4",
    pages = "044067",
    year = "2016"
}

@article{Bambi:2016BlackHole,
    author = "Bambi, Cosimo and Rubiera-Garcia, D. and Wang, Yixu",
    title = "{Black hole solutions in functional extensions of Born-Infeld gravity}",
    doi = "10.1103/PhysRevD.94.064002",
    journal = "Phys. Rev. D",
    volume = "94",
    number = "6",
    pages = "064002",
    year = "2016"
}

@article{Shaikh:2015LorentzianWormhole,
    author = "Shaikh, Rajibul",
    title = "{Lorentzian wormholes in Eddington-inspired Born-Infeld gravity}",
    doi = "10.1103/PhysRevD.92.024015",
    journal = "Phys. Rev. D",
    volume = "92",
    pages = "024015",
    year = "2015"
}

@article{Sotani:2014Properties,
    author = "Sotani, Hajime and Miyamoto, Umpei",
    title = "{Properties of an electrically charged black hole in Eddington-inspired Born-Infeld gravity}",
    doi = "10.1103/PhysRevD.90.124087",
    journal = "Phys. Rev. D",
    volume = "90",
    pages = "124087",
    year = "2014"
}

@article{Hawking:1983ThermodynamicsAdS,
    author = "Hawking, S. W. and Page, Don N.",
    title = "{Thermodynamics of Black Holes in anti-De Sitter Space}",
    reportNumber = "PRINT-83-0019 (CAMBRIDGE)",
    doi = "10.1007/BF01208266",
    journal = "Commun. Math. Phys.",
    volume = "87",
    pages = "577",
    year = "1983"
}

@article{Cognola:2005Oneloop,
    author = "Cognola, Guido and Elizalde, Emilio and Nojiri, Shin'ichi and Odintsov, Sergei D. and Zerbini, Sergio",
    title = "{One-loop f(R) gravity in de Sitter universe}",
    doi = "10.1088/1475-7516/2005/02/010",
    journal = "JCAP",
    volume = "02",
    pages = "010",
    year = "2005"
}

@article{Brevik:2004Entropy,
    author = "Brevik, Iver H. and Nojiri, Shin'ichi and Odintsov, Sergei D. and Vanzo, Luciano",
    title = "{Entropy and universality of Cardy-Verlinde formula in dark energy universe}",
    eprint = "hep-th/0401073",
    archivePrefix = "arXiv",
    doi = "10.1103/PhysRevD.70.043520",
    journal = "Phys. Rev. D",
    volume = "70",
    pages = "043520",
    year = "2004"
}

@article{Ozen:2017Entropy,
    author = "Ozen, Gokcen Deniz and Kurekci, Sahin and Tekin, Bayram",
    title = "{Entropy in Born-Infeld Gravity}",
    eprint = "1710.01110",
    archivePrefix = "arXiv",
    primaryClass = "hep-th",
    doi = "10.1103/PhysRevD.96.124038",
    journal = "Phys. Rev. D",
    volume = "96",
    number = "12",
    pages = "124038",
    year = "2017"
}

\end{document}